# Photocatalytic reduction and scavenging of Hg(II) over templated-dewetted Au on TiO$_2$ nanotubes

Davide Spanu[a,b], Alessandro Bestetti[a], Helga Hildebrand[b], Patrik Schmuki[b,c], Marco Altomare[b] and Sandro Recchia[a,*]

[a] Department of Science and High Technology, University of Insubria, via Valleggio 11, 22100 Como, Italy

[b] Department of Materials Science and Engineering WW4-LKO, University of Erlangen-Nuremberg, Martensstrasse 7, D-91058 Erlangen, Germany

[c] Chemistry Department, Faculty of Sciences, King Abdulaziz University, 80203 Jeddah, Saudi Arabia Kingdom

* Corresponding author. Email: sandro.recchia@uninsubria.it

**Abstract**

Gold-decorated $TiO_2$ nanotubes were used for the photocatalytic abatement of Hg(II) in aqueous solutions. The presence of dewetted Au nanoparticles induces a strong enhancement of photocatalytic reduction and scavenging performances, with respect to naked $TiO_2$. In the presence of chlorides, a massive formation of $Hg_2Cl_2$ nanowires, produced from Au nanoparticles, was observed using highly Au loaded photocatalysts to treat a 10 ppm Hg(II) solution. EDS and XPS confirmed the nature of the photo-produced nanowires. In the absence of chlorides and/or at lower Hg(II) starting concentrations, the scavenging of mercury proceeds through the formation of Hg-Au amalgams. Solar light driven Hg(II) abatements up to 90% were observed after 24h. ICP-MS analysis revealed that the removed Hg(II) is accumulated on the photocatalyst surface. Regeneration of Hg-loaded exhaust photocatalysts was easily performed by anodic stripping of Hg(0) and Hg(I) to Hg(II). After four catalytic-regeneration cycles only a 10% decrease of activity was observed.

**Introduction**

Pollution of water with mercury and its compounds has received increasing attention due to its toxic and bioaccumulative properties.[1–5] High levels of Hg(II) can be found in industrial wastewaters coming from activities which extensively use mercury, such as chlorine-alkali, petrochemical, metallurgical, paint and electrical industry.[1,6] Inorganic Hg(II) in water bodies can be transformed by a large number of anaerobic bacteria to methylmercury, which is a well-known neurotoxin. This compound is strongly held to fish protein when absorbed through the gills or when contaminated food sources are eaten. In some cases, methylmercury levels in carnivorous fish can be biomagnified up to a million times higher concentration levels than in the surrounding water.[5,7,8] Thus, the alkylation process of Hg(II) poses a serious health risk to humans and fauna through the aquatic food chain: even if Hg(II) is present at very low concentration, few µg L$^{-1}$ of Hg(II) could already lead to a significant mercury alkylation).

For these reasons, very restrictive worldwide regulations have been enacted with the goal of reducing mercury emissions into surface water bodies.[9] Nonetheless, the disposal of mercury currently represents a major environmental concern, and the development of new technologies for a green and efficient removal of Hg(II) from water is still a challenge.

Mercury can be successfully removed from highly concentrated aqueous solutions by membrane filtration, precipitation, ion exchange, adsorption and other methods.[10–12] However, these techniques become less efficient and more expensive with mercury concentration lower than 100 ppm.[12–14] This represents a great limitation because even in the presence of relatively low Hg concentrations high abatement rates are a fundamental prerequisite for a successful and viable removal technique.

A "green approach" that is emerging as one of the most promising method for mercury removal is the photocatalytic reduction of Hg(II), which combines simplicity of operation and could reach high performances even when treating feedstock with low mercury concentrations.[14–20]

Photocatalysis is based on the irradiation, with a proper light source, of a semiconductor, usually a metal oxide (e.g. $TiO_2$, $Fe_2O_3$, $WO_3$), which induces the promotion of electrons from the valence band (VB) to the conduction band (CB) leaving holes in the CB. The photo-generated holes and electrons can recombine (in the bulk or at the semiconductor surface) or, more preferably, can react with adsorbed species or with species in the environment to cause redox reactions.

Various metal oxide semiconductors have been explored as photocatalysts, and their field of application has been expanded in the recent years.[21–23]

Among these materials titanium dioxide ($TiO_2$) is undoubtedly considered one of the best photocatalytic material for a wide range of processes.[21,24–26] Several advantages of $TiO_2$ are its non-toxicity, low-cost, long-term stability, and corrosion and photo-corrosion stability.

In particular, nanostructured $TiO_2$, such as $TiO_2$ nanotubes (NTs),[25,27–30] having a one-dimensional (1D) morphology and a large specific surface area, exhibit a strongly enhanced photo(electro)chemical performance due to directional charge transport, together with a beneficial diffusion and adsorption geometry.[31–33]

$TiO_2$ is n-type semiconductor with a band gap of 3.00-3.20 eV, depending on its crystallographic structure. $TiO_2$ can be easily synthesized in two main crystalline forms: the most common ones are rutile (tetragonal, with a band gap of around 3.00 eV) and anatase (tetragonal, with a band gap of around 3.20 eV).[34,35]

In aqueous solution, the species that in principle can be oxidized by holes photo-generated in $TiO_2$ are water, hydroxyl ions, and, if present, organic compounds. Instead, promoted electrons can reduce oxygen (to produce superoxide radicals) or metallic ions – this given that the potential of the conduction band (CB) edge of the semiconductor is more negative than the reduction potential of the $M^{n+}/M^{(n-m)+}$ couple.[36] The knowledge of the reduction potential of the metal ions and the level of CB edge of the semiconductor is fundamental to predict the thermodynamic feasibility of the reduction: the higher the difference between these values, the higher the tendency towards reduction of metallic ions.[37] In the case of Hg(II)/Hg(0) and the CB edge of $TiO_2$, the potentials have been reported to "seat" at 0.851 V (vs NHE)[38] and ≈ - 0.50 V (at pH 7 vs NHE),[39] respectively – that is, it is thermodynamically possible to photo-reduce Hg(II) at the $TiO_2$ surface.

The photocatalytic reduction of Hg(II) over powdered $TiO_2$ has been previously reported in several works.[14–20] First attempts to photo-reduce Hg(II) on powdered $TiO_2$ photocatalysts were made by using commercial Degussa P25 and Hombikat UV100 (100% anatase) under UV light illumination.[16–18] Few years later Lenzi et al.[19] tried to improve the adsorption and photocatalytic performance by preparing via sol-gel and impregnation a $Ag/TiO_2$ powder with higher specific surface area.

To the best of our knowledge, the utilization of titania in a supported form ($WO_3$-decorated $TiO_2$ nanotubes) was reported only by Lee at al.,[14] who obtained promising mercury abatement results; ~ 76% of the initial Hg(II) content was removed in 2 hours, under illumination of a UV-B (280-315 nm, 96 W) germicidal lamp. However the utilization of such a powerful light source, and the lack of data on the total irradiance that was used, makes it difficult to compare the results with previous and future photocatalysts.

In any case, no data about the exploitation of the potential beneficial effects of the presence of gold over $TiO_2$ are reported in the literature. In principle the presence of gold should enhance the photocatalytic performances of $TiO_2$ towards Hg(II) abatement by: i) increasing the charge carriers separation and enhancing the electron transport to the $TiO_2$/electrolyte interface; ii) improving the Hg scavenging properties, through the in-situ formation of Hg-Au amalgams.

In this work we investigate for the first time, the performance of Au nanoparticles (NPs) modified $TiO_2$ nanotubes towards the solar light-driven Hg(II) photocatalytic reduction. More in detail, we found that the Hg accumulation capabilities of our $Au/TiO_2$ systems are unexpectedly high, with two main different mechanisms operating at higher or lower Hg concentrations. The influence of the presence of different amounts of gold and the regeneration/recycling ability of the photocatalysts were also explored.

**Experimental**

*Preparation procedure of Au-$TiO_2$ NTs*

Ti foils (Advent Research Materials, 0.125 mm thickness, 99.6+% purity) were degreased by sonicating in acetone, isopropanol, and deionized water and then were dried in a $N_2$ stream. The cleaned Ti foils were anodized to fabricate the highly ordered $TiO_2$ nanotube arrays at 100°C in a 3M HF solution in o-$H_3PO_4$ (Sigma- Aldrich). For this process, a two-electrode configuration was used, where the Ti foil (15 mm × 15 mm) was the working electrode and a Pt sheet was the counter electrode. The anodization was performed by applying a potential of 15 V (for 2 h) using a DC power supply (VLP 2403 Voltcraft). Subsequently, the $TiO_2$ nanotube arrays on Ti metal substrates were rinsed with ethanol and dried under $N_2$ stream.

After anodization and rinsing, an Au metal thin film was sputtered on the synthetized $TiO_2$ NTs layer by a plasma-sputtering machine (EM SCD 500, Leica) using a 99.999% pure Au target with an applied sputtering current of 16 mA and pressure of $10^{-2}$ mbar of Ar. The amount of sputtered material was in-situ determined by an automated quartz crystal monitor, and is reported in this work as nominal thickness of the sputtered film. After sputtering, the samples were annealed at 450°C in air to crystalize $TiO_2$ NTs and to induce dewetting (as described by Yoo et al.[40]). Although other deposition techniques can be used to obtain Au-decorated nanotubes (see ref.[41] as an example) we decided to use the templated-dewetting approach to obtain two morphologically different systems just changing the Au loading (i.e. just changing the sputtering time).

*Characterization techniques*

The morphology of photocatalytic material was studied using a XL30 Environmental Scanning Electron Microscopy (ESEM FEG Philips) at 20 kV under high and low vacuum conditions. Elemental analysis were performed with a silicon drifted EDS detector (EDAX element). High resolution SEM images where collected using a Hitachi S4800 FE-SEM.

XRD data were recorded on a Rigaku Miniflex system using Cu-Kα radiation, 30 kV, 10 mA with step of 0.02° (2θ) and a scanning speed of 2° min$^{-1}$.

XPS spectra were acquired using Al X-ray source. The XPS spectra were corrected in relation to the Ti2p signal at 458.5 eV. PHI MultiPak™ software and database were used for curve fitting analysis.

A Sun 2000 Solar Simulator (Abet Technologies, calibrated at 100 mW cm$^{-2}$) equipped with an AM 1.5 G filter was used as the vertical light source: a totally reflecting mirror has been used to deflect the light beam and to carry out the experiments in a side-illumination configuration.

A three-electrode system connected to an Amel 2551 potentiostat equipped with a silver chloride reference electrode (Ag/AgCl/saturated KCl), and a Pt wire as counter electrode was used for the photoelectrochemical (PEC) characterization (cyclic voltammetry and chrono-amperometry).

All measurements were performed using a 250 mL quartz beaker, placed over a magnetic stirrer: 135 g of solution were used in each test to obtain a good light focus and a total immersion of the electrodes.

*Photocatalytic Hg(II) reduction tests*

The same quartz cell and solar simulator described for the characterization techniques were used to carry out Hg(II) photo-reduction tests. Ultrapure water (MilliQ 18.2 MΩ cm, 3ppb TOC) was used for the preparation of all solutions. Abatement tests were conducted in solutions containing: i) 2 mM phosphate buffer (pH 7.0, 630 µS cm$^{-1}$) obtained by dilution of a concentrated 200 mM phosphate buffer, ii)100 mM of NaCl (prepared by dissolution of solid NaCl, Sigma-Aldrich, ≥ 98%).

The presence of Hg(II) in solution was ensured by the addition of a proper amount of a diluted solution obtained from the concentrated 1000 mg L$^{-1}$ Hg(II) solution, that was prepared by dissolving 338.4 mg of $HgCl_2$ (Carlo Erba, ≥ 99.5%) in 250 mL ultrapure water. Total mercury concentration was determined on properly diluted samples that were collected during the photocatalytic experiment with a Thermo Scientific ICAP Q inductively coupled plasma mass spectrometer (ICP-MS). The Hg concentration was determined by monitoring the 202-Hg channel and using a He-collision cell in kinetic energy discrimination (KED) mode, following the standard conditions indicated by the factory. Both external standards (prepared by dilution from a 10 mg L$^{-1}$ Hg stock solution, Merck) and internal standards (10 ppb of Germanium added in each sample by dilution from a 1000 mg L$^{-1}$ Ge stock solution, Fluka TraceSELECT®) were employed for Hg quantification. In addition, 0.2 mol L$^{-1}$ of BrCl solution, prepared by dissolving 2.7 g of KBr (Sigma-Aldrich, ≥ 99.0%) in 250 mL HCl (Fluka, > 30%, TraceSELECT®) and adding 3.8 g of $KBrO_3$ (Sigma-Aldrich, ≥ 99.0%), were added to all analyzed solutions with a 1:100 dilution, in order to minimize mercury memory effects in ICP-MS analysis.[42]

*Photoelectrochemical regeneration of Au-TiO$_2$ photocatalysts*

The same quartz cell, three-electrode system and solar simulator described for the characterization techniques were used to carry out the PEC regeneration of Au-TiO$_2$ photocatalysts. After every Hg(II) abatement experiment, Au-TiO$_2$ NTs samples were dipped in a 100 mM KNO$_3$ (prepared by dissolving solid KNO$_3$, Sigma-Aldrich, ≥ 99.0%) solution and a positive +500 mV (vs. Ag/AgCl) potential was applied for 2 hours under stirring and front solar-irradiation. Cyclic voltammograms from -200 mV to +1000 mV (vs. Ag/AgCl) with a scan speed of 200 mV s$^{-1}$ were recorded in this solution under permanent illumination conditions. The resulting solutions were analyzed in view of their composition with ICP-MS as described for the samples collected for photocatalytic reduction.

**Results and discussion**

Fig. 1a shows the morphology of the top and cross-sectional view (see inset in Fig. 1a) of the highly-ordered $TiO_2$ NTs employed in this work, which are anodically grown on Ti metal foils in a hot, concentrated $H_3PO_4$/HF electrolyte.[40] These well-defined nanostructures exhibit a nearly ideal hexagonal packing and each tube has an average internal diameter of ~ 90 nm and a length of ~ 200 nm.

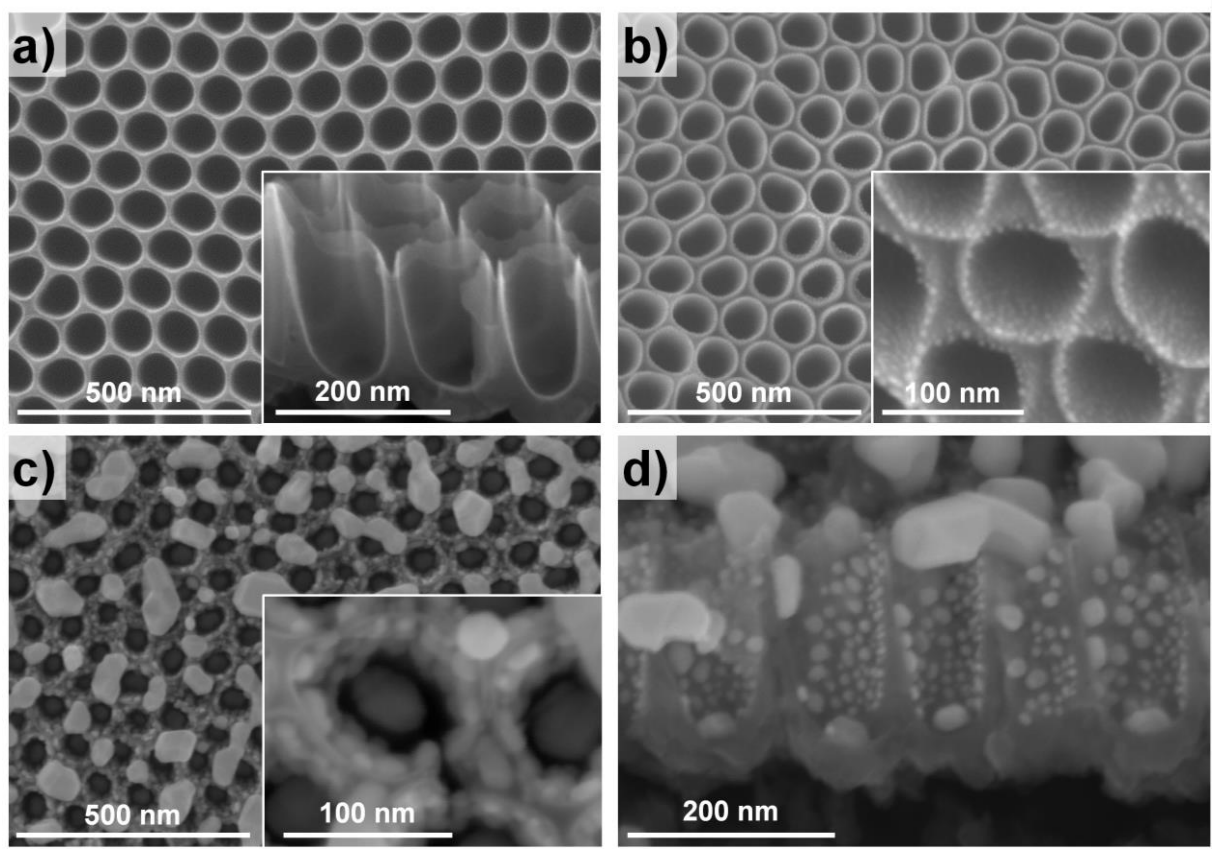

Fig. 1. (a-d) SEM images of $TiO_2$ structures employed in this work: (a) top view of pristine $TiO_2$ nanotubes. Inset: cross-sectional view of pristine $TiO_2$; (b) $TiO_2$ NTs coated with a sputtered 1 nm-thick Au film after dewetting (1Au-$TiO_2$). Inset: magnified top view image of 1Au-$TiO_2$; (c) $TiO_2$ NTs coated with a sputtered 20 nm-thick Au film after dewetting (20Au-$TiO_2$). Inset: magnified top view image of 20Au-$TiO_2$; (d) cross-sectional view of 20Au-$TiO_2$

As reported in previous work these NT arrays can be decorated with nanoscopic precision by sputter-deposition of a thin metal film (Au in this work) followed by a proper thermal treatment that induces both the crystallization of $TiO_2$ and the solid-state templated-dewetting of the metal.[40,43–48] Images of TiO$_2$ NTs coated with a nominally 1 nm-thick (1Au-TiO$_2$) and 20 nm-thick (20Au-TiO$_2$) Au layers after a heat treatment (i.e. after dewetting) are reported in Fig. 1b and 1c-d, respectively. Thermal dewetting of metal thin films occurs at the solid-state: the temperature of the thermal treatment is far below the melting point of Au. Essentially, the driving force for dewetting is related to the minimization of: i) the free surface energy of the oxide surface (TiO$_2$), ii) the free surface energy of the metal layer (Au) and iii) the interfacial energy of the metal/substrate system (Au/TiO$_2$).[45,49]

For the sake of comparison, Fig. S1a and S1b show the SEM images recorded prior dewetting for 1Au-TiO$_2$ and 20Au-TiO$_2$, respectively: in the case of the 20 nm deposition a strong difference between pristine and dewetted samples can be observed, while for the 1 nm specimen no significant differences can be appreciated, as for this sample dewetting occurs at nearly room temperature (so thin layers require low energy input to dewet).[45] Analogous structures (same anodization conditions, thermal treatment and sputtered metal) were prepared and characterized (XRD, XPS and hydrogen evolution capability under UV irradiation) in a previous work.[40]

In this work, Au-TiO$_2$ systems were assessed in view of their photocatalytic Hg(II) reduction performance under solar light simulated irradiation. Both 1Au-TiO$_2$ and 20Au-TiO$_2$ were tested over solutions buffered at pH 7.0 (see experimental section), having two markedly different Hg starting concentrations.

Fig. 2 shows the results of the photocatalytic abatement tests of both 1Au-TiO$_2$ and 20Au-TiO$_2$ systems, carried out with solutions having a starting Hg concentration of 10 ppm (Fig. 2a) or 500 ppb (Fig. 2b).

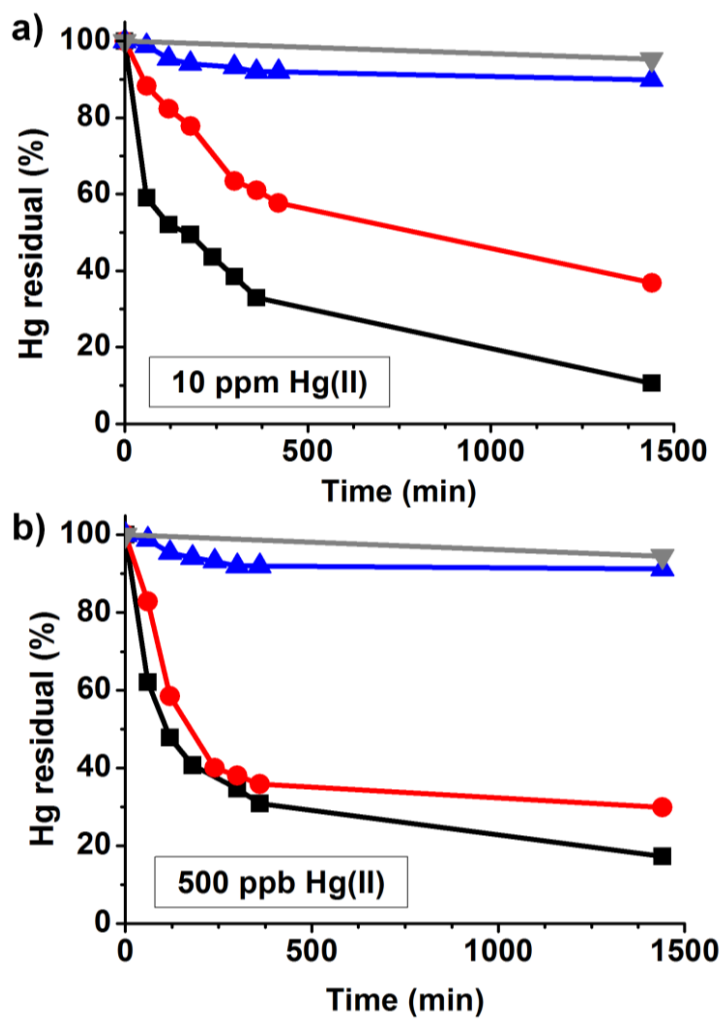

**Fig. 2. (a-b)** Hg(II) concentration trends during photocatalytic reduction using 20Au-TiO$_2$ (black), 1Au-TiO$_2$ (red), TiO$_2$ (blue) and 20Au-TiO$_2$ in dark condition (grey) at pH 7.0 in presence of chloride with different starting Hg(II) concentrations: (a) 10 ppm and (b) 500 ppb.

The reported Hg reduction profiles are surely correlated to photo-activated processes, as no significant Hg abatement is observed in the dark. Moreover, a comparison of the abatement profiles obtained for naked TiO$_2$ and for both Au-TiO$_2$ systems strongly indicates that gold decoration is fundamental to obtain high photocatalytic Hg(II) reduction (removal) rates. As a preliminary consideration, the strong promotion effect caused by the presence of Au NPs might be ascribed to: i) the improvement of the electron transfer to reactants provided by Au NPs (localized Schottky-junctions);[28] ii) the strong interaction between Hg and Au that can lead to the formation of nano-alloys ("nano-amalgams", if Hg(II) is completely reduced to Hg(0) and incorporated in the Au NPs). In particular, the best photocatalytic performance was obtained with the most Au-loaded photocatalyst (20Au-TiO$_2$), which is able to reduce up to ~ 90 % of Hg(II) after 24 hours, almost independently of the starting Hg(II) concentration. Instead, 1Au-TiO$_2$ exhibits a certain sensitivity to the mercury starting concentration: the higher the Hg(II) starting concentration, the lower the photocatalytic efficiency.

During the abatement of the 10ppm Hg solution the formation of a compact white-grayish layer was macroscopically observed on 20Au-TiO$_2$. Surprisingly, SEM images recorded on this specimen at the end of the abatement test revealed that the catalysts' surface is homogeneously covered by nano filaments, having a ~ 100 nm mean width and length of some tens of microns (Fig. 3a). The formation of such nanofibers is very reproducible, as it was always observed with different 20Au-TiO$_2$ specimens and even after each regeneration and recycling test (discussed below). No detachment of the compact layer of such nanowires from the catalyst surface was observed even after 24h. EDS analysis carried out on several of these filaments revealed that they are mainly composed by Hg and Cl, with an approximate Hg/Cl atomic ratio of 1.1. It therefore reasonable to suggest the formation of insoluble Hg$_2$Cl$_2$ filaments, which is possible due to the presence of NaCl in the aqueous medium (additional characterization data to support the formation of Hg$_2$Cl$_2$ are shown below).

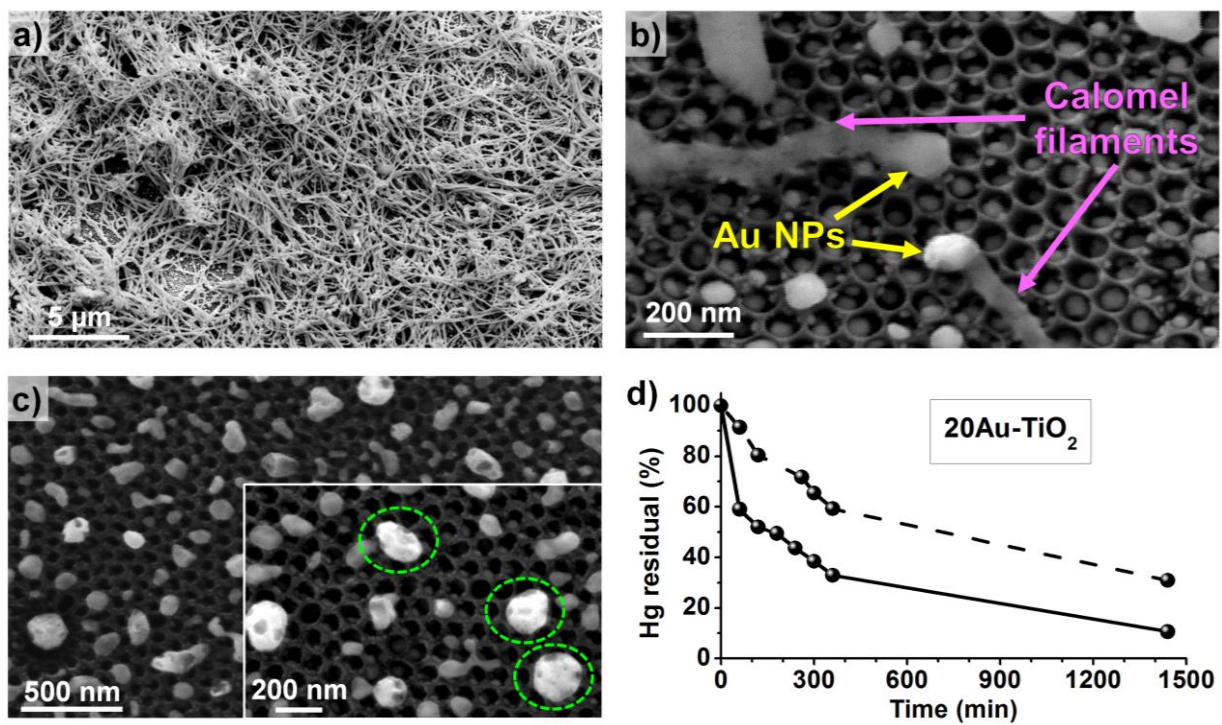

**Fig. 3.** (a-b) SEM images of 20Au-TiO$_2$ recorded after (a) 24 hours and (b) 30 minutes long photocatalytic reduction of 10 ppm of Hg(II) in presence of chloride. (c) 20Au-TiO$_2$ after 24 hours long photocatalytic reduction of 10 ppm of Hg(II) in absence of chloride. Inset: magnified image in which larger and brighter Au NPs are circled in green. (d) Hg(II) concentration trends during photocatalytic reduction of 10 ppm of Hg(II) at pH 7.0 in presence (solid line) and absence (dashed line) of chlorides using 20Au-TiO$_2$.

NaCl was intentionally introduced in solution for two main reasons: i) chlorides are always present in every Hg polluted aqueous medium (ranging from fresh to marine waters): ii) it was reported by Wang et al.[16] that with powdered TiO$_2$ photocatalysts the presence of Cl$^-$ is detrimental in terms of Hg(II) reduction capability, due to the formation of a variety of complexes, including HgCl$^+$, HgCl$_3^-$ and HgCl$_4^{2-}$.

SEM images taken form a 20Au-TiO$_2$ specimen after an intentionally short reduction test (Fig. 3b) clearly revealed that Au NPs act as nucleation sites for the directional growth of calomel nanowires. This evidence, together with the observation that the diameters of calomel nanowires (as measured from Fig. 3a) are slightly lower than the diameters of Au NPs, would indicate that these wires exclusively grow starting from Au NPs.

It should be underlined that the formation of calomel nanowires is relevant as such because it strongly enhances the scavenging properties of the Au-TiO$_2$ photocatalyst (this is key especially when highly concentrated Hg solutions have to be treated). In the absence of NaCl, the formation of Hg$_2$Cl$_2$ filaments was not observed (Fig. 3c), and in these conditions the photocatalytic Hg reduction is significantly slower (Fig. 3d). We report in Fig. 3c the SEM image of sample 20Au-TiO$_2$ after the reduction tests carried out in the absence of chlorides: the presence of some larger and brighter metallic NPs, which were not present before the photocatalytic experiment (Fig. 1c), is clearly visible. EDS analysis revealed that these larger NPs are composed of Au and Hg with a variable atomic Hg-Au ratio, which ranges from ~ 0.8 to 1.8. These results are consistent with the formation of an amalgam between Hg(0) and Au, and in this case the mechanism would probably involve the complete reduction of Hg(II) to Hg(0).

XRD patterns recorded prior and after the photocatalytic test carried out without chlorides (compare patterns a-d in Fig. 4) show a substantial loss of crystallinity of the Au NPs: the decrease in intensity of Au (200) reflection at ~ 44.6° is consistent with the suggested formation of Au-Hg alloy.[50,51] As we will show later, this is also confirmed by XPS analysis. A partial loss of crystallinity was observed also in the XRD spectrum (Fig. 4b) recorded after the photoreduction test carried out in the presence of chlorides (i.e. with the formation of Hg$_2$Cl$_2$ filaments): therefore, the partial formation of Hg-Au amalgams may not be excluded even in this case; however, owing to the structure coverage caused by the calomel nanowires, we could not observe by SEM the formation of structures comparable to those in Fig. 3d (Hg-Au NPs).

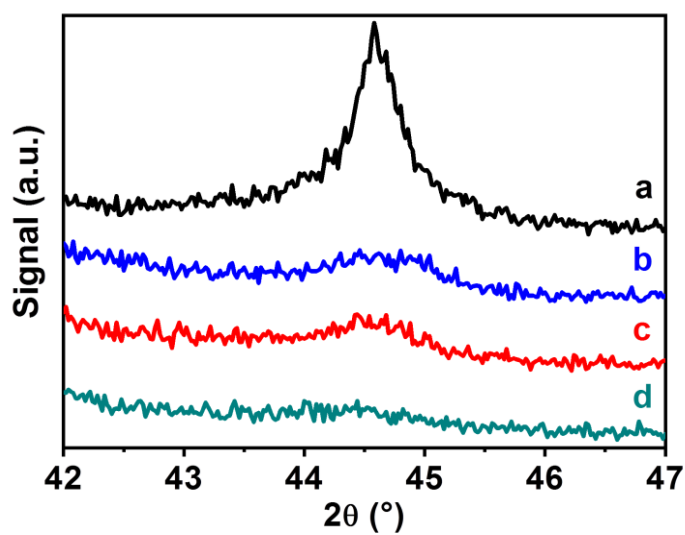

**Fig. 4.** Detail of Au (200) reflection in XRD spectra for 4 differently treated samples: (a) as-prepared 20Au-TiO$_2$; (b) 20Au-TiO$_2$ after photocatalytic reduction of 10 ppm of Hg(II) in presence of chloride; (c) the same material reported in (b) after PEC regeneration in KNO$_3$; (d) 20Au-TiO$_2$ after photocatalytic reduction of 10 ppm of Hg(II) in absence of chloride. Full XRD spectra are provided in Fig. S2.

The XPS spectra reported in Figs 5a-d and Fig. S3 further confirm the formation of calomel nanowires in the presence of chlorides, and the formation of Hg-Au alloy in the absence of chloride. The XPS analysis of 20Au-TiO$_2$ after photoreduction of Hg(II) in the presence of chlorides revealed two peaks at 101.05 eV and 105.10 eV which can be attributed to the Hg4f$_{7/2}$ and Hg4f$_{5/2}$ signals of Hg$_2$Cl$_2$.[52,53] Instead, the formation of Hg-Au alloy (i.e. Hg(0)) in the absence of chlorides is supported by a shift and be the broadening of the Hg signals, which can fitted according to two doublets peaking at 98.90 and 102.79 eV, and 99.99 and 103.92 eV (Hg4f$_{7/2}$ and Hg4f$_{5/2}$, respectively).[53,54] The doublet nature and broadening of the Hg XPS signal reflects that Hg(0) is present with different chemical surroundings; such signal can e.g. be ascribed to the formation of Hg-Au alloy with various stoichiometric composition,[55] and to the presence of "free" Hg(0) adsorbed on TiO$_2$ NTs or at the Au NP surface.[54] As shown in Fig. S3, the Au4f XPS doublet for as-prepared 20Au-TiO$_2$ peaks at 83.10 eV and 86.70 eV, which corresponds well to data in the literature on Au deposited on TiO$_2$.[40] These signals however seem not to shift with the formation of calomel or Hg-Au alloy; this results is also in line with data in the literature.[54]

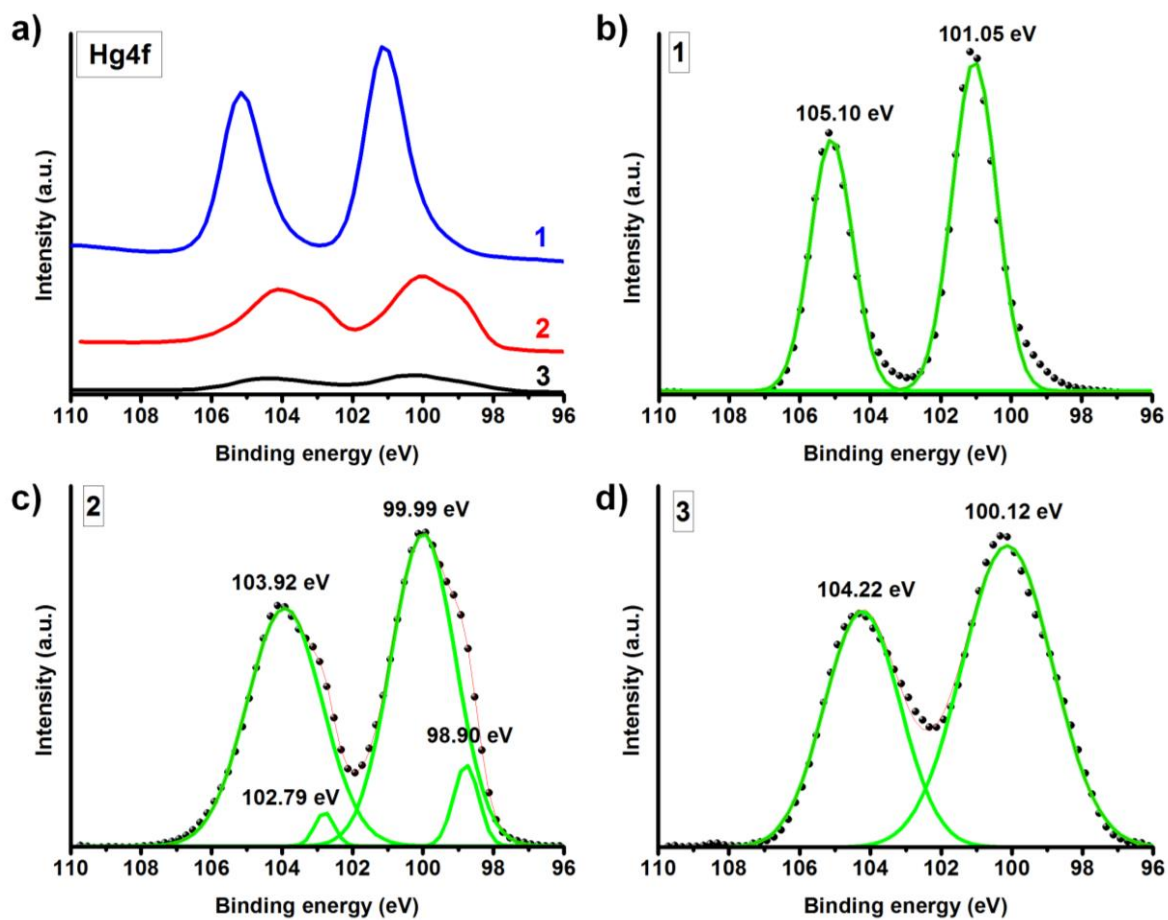

**Fig. 5. (a) Comparison between Hg4f XPS spectra of (1) 20Au-TiO$_2$ after photocatalytic reduction of 10 ppm of Hg(II) in presence of chlorides, (2) 20Au-TiO$_2$ after photocatalytic reduction of 10 ppm of Hg(II) in absence of chlorides, (3) 20Au-TiO$_2$ after PEC regeneration. (b-d) Deconvolution of Hg4f signals for samples 1-3.**

All the results so far reported can be rationalized as follows: i) Hg(II) → Hg(0) proceeds stepwise with the intermediate formation of Hg(I); ii) in the presence of Cl$^-$ the concentration of Hg(I) produced by photoreduction in close proximity of Au NPs is sufficiently high to grow Hg$_2$Cl$_2$ nanowires starting from such NPs; iii) in the absence of Cl$^-$, the reduction further proceeds to Hg(0), i.e. with the formation of Hg-Au nano-amalgam; iv) a partial formation of Hg-Au amalgams cannot be excluded even in the presence of chlorides.

Apart from the presence of chlorides, the formation of calomel nanowires strongly depends on the amount of gold deposited on $TiO_2$ NTs and on the initial Hg concentration. In fact, the formation of such nanowires was neither observed on 1Au-$TiO_2$ (independently of the initial Hg concentration), nor on 20Au-$TiO_2$ for Hg concentration > 500 ppb (see Fig. S4a and S4b, respectively). Based on these results, we can infer that sufficiently big Au NPs, together with high Hg(II) concentrations, are fundamental to generate a sufficiently high Hg(I) local concentration to produce calomel nanowires. Apparently, the mechanism for such a directional and anisotropic growth of calomel nanowires roughly resembles the growth of carbon nanotubes starting from metal NPs,[56,57] with the obvious difference that here we are working in the liquid phase rather than in the gas phase.

In any case, when the conditions are suitable for the growth of calomel nanowires, no $Hg_2Cl_2$ filaments produced from Au NPs inside $TiO_2$ nanotubes were observed, as well as from $TiO_2$. Additionally, when the conditions are not suitable for the growth of calomel nanowires, adsorption of Hg over $TiO_2$ surely occurs, as already reported in the literature.[14,58,59] This evidence is only indirect (EDS resolution is not sufficient to locate Hg on $TiO_2$ rather than on Au), since the Au loading on $TiO_2$, cannot account for the scavenged Hg amounts.

Another investigated aspect was the possibility of reusing the Au-$TiO_2$ photocatalysts after a proper regeneration process. This investigation was conducted on 20Au-$TiO_2$ only, due to its superior performances at high Hg concentrations.

Since photo-reduced Hg is mainly loaded on Au NPs and on calomel nanowires bound to the Au NPs, the most rational way to efficiently reuse the photocatalyst is to selectively dissolve mercury, leaving behind Au NPs on $TiO_2$ NTs, that is, regenerating the original photocatalyst morphology and composition. In principle this process can be performed trough an electrochemical oxidation of the Hg-loaded photocatalyst, in a similar way to the anodic stripping determination of Hg on gold electrodes. This process would allow i) the regeneration of the photocatalyst (Hg-free Au NPs), and ii) the recovery of mercury in a small concentrated waste volume.

Therefore we investigated the feasibility of using an electrochemical dissolution in a $KNO_3$ 100 mM media to regenerate the electrode. $KNO_3$ was chosen because of the high solubility of both mercuric and mercurous nitrates.[60] It should be underlined that, since $TiO_2$ is a semiconductor, it is mandatory to perform electrochemical dissolution under light illumination to produce the necessary charge carriers (as a matter of fact, any attempt to regenerate Au-$TiO_2$ catalysts applying up to 800 mV in the dark failed). In order to establish the regeneration potential to be applied to the Hg-loaded photocatalyst, cyclic voltammograms (CVs) were recorded in $KNO_3$ under illumination (Fig. 6a). One can observe the anodic region to start slightly above 0 mV (vs. Ag/AgCl). Moreover, as the stripping potential of Hg over gold electrodes is reported to fall on around 350-450 mV[61,62] we decided to use 500 mV (a slightly more positive potential) as the regeneration potential.

From the current vs. time plot (Fig. 6b) recorded during the PEC regeneration it is possible to observe that a constant current value is reached after 120 min. The regeneration time was accordingly set to 120 min.

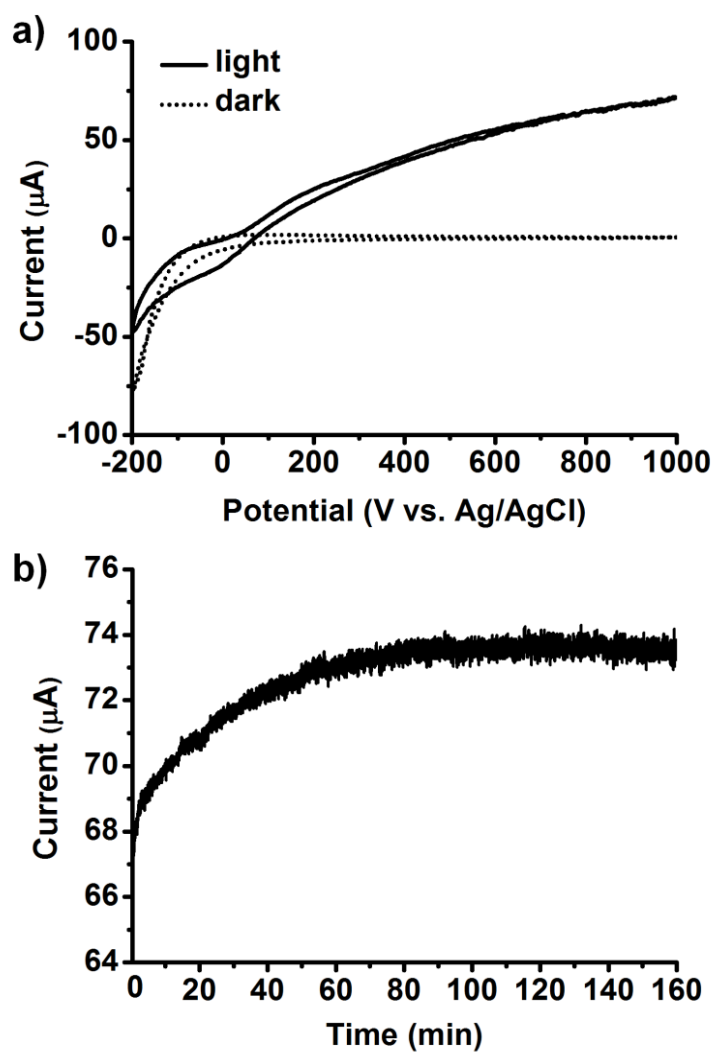

**Fig. 6.** (a) Cyclic voltammograms recorded using 20Au-TiO$_2$ in KNO$_3$ 100 mM with and without solar light irradiation; (b) chronoamperometric measurements recorded during 20Au-TiO$_2$ regeneration in 100mM KNO$_3$ solution applying 500 mV (vs. Ag/AgCl).

EDS analysis performed after PEC regeneration on different 20Au-TiO$_2$ samples revealed that the treatment was very successful: in the worst case only a ~ 0.14 Hg/Au atomic ratio was determined (please note that the Hg/Au ratios after Hg abatement were ~ 0.8-1.8). Furthermore, ICP-MS analysis of the KNO$_3$ solution after regeneration revealed the presence of more than 80 % of the previously photocatalytically removed Hg(II). These results fit very well with the XPS analysis of 20Au-TiO$_2$ after PEC regeneration (Fig. 5): it is possible to observe a strong decrease of the XPS signals at 100.12 eV and 104.22 eV, which, as outlined above, is attributed to Hg(0).[53,54]

As shown in Fig. 7, the regeneration process does not cause any modification of the Au NPs shape, and leads to a complete cleaning of the Au-TiO$_2$ structures from calomel nanowires. The only relevant difference between fresh and regenerated photocatalysts is reported in Fig. 4c, where it can be seen that the loss of crystallinity of Au NPs observed during the catalytic test is not recovered after regeneration.

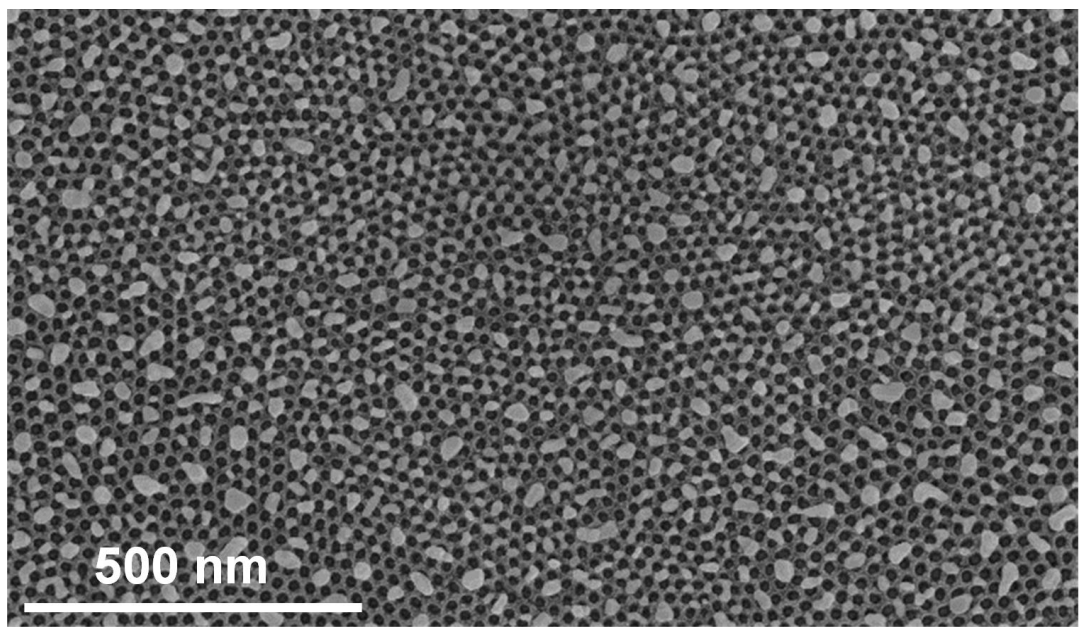

**Fig. 7. SEM image of 20Au-TiO$_2$ after PEC regeneration procedure.**

The abatement performances obtained for a freshly prepared 20Au-TiO$_2$ photocatalyst and for three subsequent recycling tests (a regeneration cycle was applied between each test) are schematically reported in Fig. 8. As it can be seen, the loss of activity is quite low even after 4 catalytic runs, indicating thus the effectiveness of the applied regeneration approach. These results are even more attractive if we consider that regeneration parameters were not optimized (this will be subject to further work).

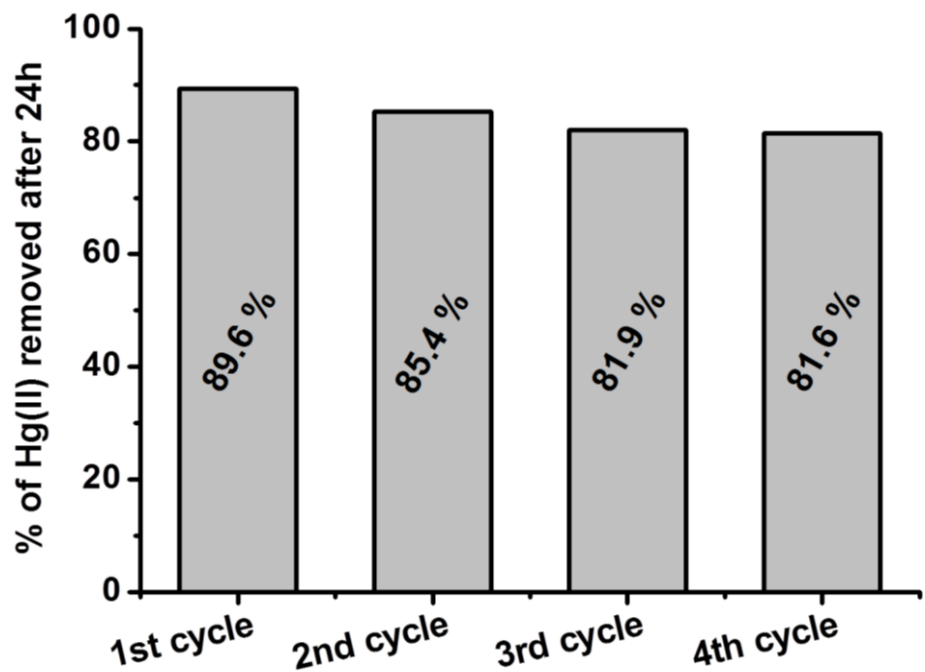

**Fig. 8.** Percentage of Hg(II) removed from 10 ppm Hg(II) chlorides-containing solutions after 24 hours long photocatalytic experiments carried out using 20Au-TiO$_2$ re-cycled up to 4 times

**Conclusions**

In this paper we have shown that Au-TiO$_2$ photocatalysts are able under sun light irradiation to efficiently photo-reduce Hg(II) to Hg(I) or Hg(0) and to accumulate these species. This functionality is enabled by the Au NPs at TiO$_2$ NT surface.

Two different mechanisms can be identified depending on the chloride concentration, Hg(II) concentration and Au NPs size. The first involves the partial reduction of Hg(II) to Hg(I) and, as a result, produce the massive formation of insoluble calomel nanowires that occurs when 20Au-TiO$_2$ is used to treat solution of quite high Hg(II) concentration levels and containing chlorides. A second mechanism involves the complete reduction of Hg(II) to Hg(0) and the formation of Hg-Au amalgam; this seems to occur whenever it is not possible to form insoluble Hg$_2$Cl$_2$.

Best results, in terms of Hg abatement, were obtained with the most Au-loaded photocatalyst (20Au-TiO$_2$). It is noteworthy that this photocatalyst has an intrinsic bimodal abatement behavior: at higher Hg concentrations the abatement proceeds through the formation of calomel wire (with attracting scavenging features), while the formation of Hg-Au amalgam is preferred at low Hg concentrations.

We also demonstrated the feasibility of an efficient regeneration of the photocatalyst by PEC anodic stripping, which additionally allows to recover the previously removed Hg(II) in a concentrated waste.

**Acknowledgements**

The authors would like to acknowledge ERC (340511), DFG and the DFG cluster of excellence EAM (EXC 315) for financial support. D.S. and S.R. gratefully acknowledge financial support from MIUR.